\documentclass[11pt]{article}
\usepackage{epsfig}
\usepackage{amsfonts,latexsym}

\catcode`\@=11
\def\Let@{\relax\iffalse{\fi\let\\=\cr\iffalse}\fi}
\def\vspace@{\def\vspace##1{\crcr\noalign{\vskip##1\relax}}}
\def\multilimits@{\bgroup\vspace@\Let@
 \baselineskip\fontdimen10 \scriptfont\tw@
 \advance\baselineskip\fontdimen12 \scriptfont\tw@
 \lineskip\thr@@\fontdimen8 \scriptfont\thr@@
 \lineskiplimit\lineskip
 \vbox\bgroup\ialign\bgroup\hfil$\m@th\scriptstyle{##}$\hfil\crcr}
\def\Sb{_\multilimits@}
\def\endSb{\crcr\egroup\egroup\egroup}
\def\Sp{^\multilimits@}

\catcode`\@=11

\def\vereq#1#2{\lower3pt\vbox{\baselineskip1.5pt \lineskip1.5pt
\ialign{$\m@th#1\hfill##\hfil$\crcr#2\crcr\sim\crcr}}}

\newcommand{\be}[1]{\begin{equation}\label{#1}}
\newcommand{\ee}{\end{equation}}
\newcommand{\ba}[1]{\begin{eqnarray}\label{#1}}
\newcommand{\ea}{\end{eqnarray}}

\newcommand{\bmatrix}[1]{\left( \begin{array}{#1}}
\newcommand{\ematrix}{\end{array}\right)}

\newlength{\indentedwidth}
\newdimen\mathindent
\indentedwidth=\mathindent

\newenvironment{indented}{\begin{indented}}{\end{indented}}
\newenvironment{varindent}[1]{\begin{varindent}{#1}}{\end{varindent}}
\def\indented{\list{}{\itemsep=0\p@\labelsep=0\p@\itemindent=0\p@
   \labelwidth=0\p@\leftmargin=\mathindent\topsep=0\p@\partopsep=0\p@
   \parsep=0\p@\listparindent=15\p@}\footnotesize\rm}

\def\varindent#1{\setlength{\varind}{#1}%
   \list{}{\itemsep=0\p@\labelsep=0\p@\itemindent=0\p@
   \labelwidth=0\p@\leftmargin=\varind\topsep=0\p@\partopsep=0\p@
   \parsep=0\p@\listparindent=15\p@}\footnotesize\rm}

\begin{document}
%%%%%%%%%%%%%%%%%%%%%%%%%%%%%%%%%%%%%%%%%%%%%%%%%%%%%%%%%%%%%%%%%%%%%%%%%%%%%%%%%%%
\smallskip
\smallskip
\author{Alexey A. Kryukov \\
%{\small Department of Mathematics}\\
{University of Wisconsin}\\
%{\small 5119 Helen C. White, 600 N. Park St., Madison, WI 53706}\\
%{\small Department of Mathematics,}\\
%{\ University of Wisconsin}\\
%\\
\smallskip
{\small e-mail: aakrioukov@facstaff.wisc.edu}\\
\\
\smallskip
%{\small Key words: Hilbert manifolds, generalized functions, partial differential equations}\\
%\smallskip
%{\small PACS: 02.40.Ky, 02.40.Tt, 03.30.+p, 04.50.+h}\\
}

\title{Coordinate formalism on abstract Hilbert space}

\maketitle

\begin{abstract}
Coordinate formalism on Hilbert manifolds developed in \cite{Kryukov} is reviewed.
The results of \cite{Kryukov} are applied to the simpliest case of a Hilbert manifold: 
the abstract Hilbert space. In particular, functional transformations preserving properties 
of various linear operators on Hilbert spaces are found. Any generalized solution of an
arbitrary linear differential equation with constant coefficients is shown to be related 
to a smooth solution by a (functional) coordinate transformation.
The results also suggest a way of using generalized functions to solve nonlinear
problems on Hilbert spaces.
  
\end{abstract}

\bigskip

%%%%%%%%%%%%%%%%%%%%%%%%%%%%%%%%%%%%%%%%%%%%%%%%%%%%%%%%%%%%%%%%%
%
%\newpage

\section{Introduction}

\setcounter{equation}{0}

In \cite{Kryukov} a coordinate formalism on abstract infinite-dimensional
Hilbert manifolds has been
introduced. By letting images of charts on a Hilbert manifold belong to
arbitrary Hilbert spaces of functions we were able to find the new
infinite-dimensional counterparts of notions of a basis, dual basis,
orthogonal basis etc. We found that the choice of a functional Hilbert model for a 
Hilbert manifold (rather than the choice of a chart within a given model)
is deeply similar to the choice of coordinates on a finite dimensional
manifold.

In this paper we continue investigating the obtained formalism. The main attention
here is on the simpliest case of a Hilbert manifold: the abstract Hilbert space.

In section 2 the main definitions of \cite{Kryukov} are briefly reviewed.

In section 3 we consider the coordinate transformations preserving locality of a
given operator. As a particular solution of the locality equations we obtain 
the Fourier transform.

In section 4 the coordinate transformations preserving the derivative operator 
are investigated.
It is shown in particular that the generalized and the smooth solutions
of linear differential equations with constant coefficients are related by a
change of functional coordinates.

In section 5 we analyze the coordinate transformations preserving the operator 
of multiplication by a function. It is shown that except for the trivial transformations
locality of the operator of multiplication by a function can not be preserved.

In section 6 more general coordinate transformations are considered. In particular, we show
how the results of section 4 can be generalized to the case of linear differential 
equations with the non-constant coefficients. We also start investigating here
the case of nonlinear differential equations.

The results are briefly summarized in the conclusion.

\section{Linear algebra on the string space}

\setcounter{equation}{0}

{\it Definition}. The {\it string space} $\bf {S}$ is an abstract vector space
that is also a differentiable manifold linearly diffeomorphic to an 
infinite-dimensional separable Hilbert space.

{\it Definition}.  A {\it Hilbert space of functions} is either a Hilbert space 
$H$, elements of which are equivalence classes of maps between two given
subsets of $R^{n}$ or the Hilbert space $H^{\ast }$ dual to $H$. Two
elements $f,g\in H$ are called {\it equivalent} if the norm of $f-g$ in $H$ is
zero.

{\it Definition}.  A linear isomorphism ${e}_{H}$ from a Hilbert space $H$ 
of functions onto $\bf {S}$ 
will be called a {\it string basis} on $\bf {S}$.

{\it Notation}. The action of ${e}_{H}$
on $\varphi \in H$ will be written in one of the following ways: 
\begin{equation}
{e}_{H}(\varphi )=({e}_{H},\varphi )=\int {e}_{H}(k)\varphi (k)dk={e}%
_{Hk}\varphi ^{k}.
\end{equation}
The integral sign is used as a notation for the action of ${e}_{H}$ on an
element of $H$ and in general does not refer to an actual integration. 

{\it Definition}. Given ${e}_{H}$ the function $\varphi \in H$ such that $\Phi={e}_{H}\varphi$
will be called a {\it coordinate} (or an {\it $H$-coordinate}) of a string $\Phi \in {\bf {S}}$.
The space $H$ itself will be called a {\it coordinate space}.

{\it Definition}. Let ${\bf {S}}^{\ast}$ be the dual string space. 
A linear isomorphism ${e}_{H^{\ast }}$ of $H^{\ast }$ onto ${\bf {S}}^{\ast}$ 
will be called a {\it string basis on} ${\bf {S}}^{\ast}$.

{\it Notation}. Decomposition of an
element ${F}\in \bf{{S}^{\ast }}$ with respect to the basis will be
written in one of the following ways:
\begin{equation}
F={e}_{H^{\ast }}(f)=({e}_{H^{\ast }},f)=\int {e}_{H^{\ast }}(k)f(k)dk=
{e}_{H^{\ast }}^{k}f_{k}.
\end{equation}

{\it Definition.} The basis ${e}_{H^{\ast }}$ will be called {\it dual} to the
basis ${e}_{H}$ if for any string ${\Phi }={e}_{Hk}\varphi ^{k}$ and for any
functional ${F}={e}_{H^{\ast }}^{k}f_{k}$ the following is true: ${F}({\Phi })=f(\varphi )$.

{\it Remark}. In general case we have 
\begin{equation}
F(\Phi )=e_{H^{\ast }}f(e_{H}\varphi )=e_{H}^{\ast }e_{H^{\ast }}f(\varphi ),
\end{equation}
where $e_{H}^{\ast }:{\bf {S}}^{\ast }\longrightarrow H^{\ast }$ is the
adjoint of $e_{H}$. Therefore, $e_{H^{\ast }}$ is the dual string basis if $%
e_{H}^{\ast }e_{H^{\ast }}:H^{\ast }\longrightarrow H^{\ast }$ is the
identity operator. 

{\it Notation}. The action of $F$ on $\Phi $ in any bases $e_{H}$ on $\bf {S}$ and $e_{H^{\ast}}$ 
on ${\bf {S}}^{\ast}$ will be written in one of the following ways: 
\begin{equation}
F(\Phi )=e_{H^{\ast }}^{k}f_{k}e_{Hl}\varphi ^{l}=G(f,\varphi
)=g_{l}^{k}f_{k}\varphi ^{l},
\end{equation}
where $G$ is a non-degenerate bilinear functional on $H^{\ast }\times H$.

{\it Notation}. Assume that $H$ is a real Hilbert space. We have:
\begin{equation}
\label{orto}
(\Phi ,\Psi )_{S}={\bf{G}}(\Phi ,\Psi )=G(\varphi,\psi 
)=g_{kl}\varphi ^{k}\psi ^{l}. 
\end{equation}
Here $G:H\times H\longrightarrow R$ is a bilinear form defining the inner
product on $H$ and ${\bf {G}}:{\bf {S}}\times {\bf {S}}\longrightarrow R$
is the induced bilinear form. The expression on the right is a convenient
form of writing the action of $G$ on $H\times H$. 

{\it Definition}.  A string basis ${e}_{H}$ in $\bf {S}$ will be called
{\it orthonormal} if 
\begin{equation}
(\Phi ,\Psi )_{S}=f_{\varphi}(\psi ),
\end{equation}
where $f_{\varphi}=(\varphi,\cdot)$ is a regular functional and $\Phi=e_{H}\varphi$,
$\Psi=e_{H}\psi$ as before. That is, 
\begin{equation}
\label{orto2}
(\Phi,\Psi)=f_{\varphi}(\psi)=\int \varphi(x)\psi(x)d\mu(x),
\end{equation}
where $\int$ here denotes an actual integral over a $\mu$-measurable set $D \in R^{n}$.

{\it Remark}. Not every coordinate Hilbert space $H$ can
produce an orthonormal string basis ${e}_{H}$.
Equation (\ref{orto2}) shows that orthonormality of a string basis imposes a 
symmetry between coordinates of the
dual objects in the basis. In particular, if $e_{H}$ is
orthonormal, then $H$ must be an $L_{2}$-space, i.e. a space $L_{2}(D,\mu)$
of square integrable functions on a $\mu$-measurable set $D \in R^{n}$. 
Thus, Hilbert spaces $l_{2}$ and $L_{2}(R)$ are examples of
coordinate spaces that admit an orthonormal string basis.

{\it Definition.} A linear coordinate transformation on $\bf {S}$ is 
an isomorphism $\omega :\widetilde{H}\longrightarrow H$ of Hilbert 
spaces which defines a new string
basis $e_{\widetilde{H}}:\widetilde{H}\longrightarrow \bf {S}$ by 
$e_{\widetilde{H}}=e_{H}\circ \omega $. 

Let $\varphi $ be coordinate of a string $
\Phi $ in the basis $e_{H}$ and $\widetilde{\varphi }$ its coordinate in the
basis $e_{\widetilde{H}}$. Then $\Phi =e_{H}\varphi =e_{\widetilde{H}}
\widetilde{\varphi }=e_{H}{\omega} \widetilde{\varphi }$. That is, $\varphi
=\omega \widetilde{\varphi }$ by the uniqueness of the decomposition. 

Let now $\Phi,\Psi \in {\bf {S}}$ and let ${\bf {A}}$ be a linear operator on $\bf {S}$. 
Let $\Phi=e_{H}\varphi, \Psi=e_{H}\psi$ with $\varphi,\psi \in H$.
The scalar product $(\Phi,{\bf {A}}\Psi)_{S}$ is independent of a basis 
and in a basis $e_{H}$ reduces to 
\begin{equation}
(\varphi,A\psi)_{H}=(\widehat{G}\varphi,A\psi), 
\end{equation}
where $\widehat{G}:H \longrightarrow H^{*}$ defines
the metric on $H$. 

If $\omega: \widetilde{H}\longrightarrow H$ is a linear coordinate
transformation and $\varphi=\omega\widetilde{\varphi}, \psi=\omega\widetilde{\psi}$, then
\begin{equation}
(\widehat{G}\varphi,A\psi)=(\widehat{G}\omega\widetilde{\varphi},A\widetilde{\psi})=
(\omega^{*}\widehat{G}\omega\widetilde{\varphi},\omega^{-1}A\omega\widetilde{\psi})=
(\widehat{G}_{\widetilde{H}}\widetilde{\varphi},A_{\widetilde{H}}\widetilde{\psi}).
\end{equation} 
Therefore we have the following transformation laws:
\begin{eqnarray}
\varphi &=&\omega\widetilde{\varphi}\\ 
\psi &=&\omega\widetilde{\psi}\\
\widehat{G}_{\widetilde{H}} &=&\omega^{*}\widehat{G}\omega\\
A_{\widetilde{H}} &=&\omega^{-1}A\omega.
\end{eqnarray}

More generally, we know that by definition $\bf {S}$ is a Hilbert manifold (modelled on itself). 
Let then $(U_{\alpha },\pi _{\alpha })$ be an atlas on $\bf {S}$. 

{\it Definition}. A collection of quadruples $(U_{\alpha },\pi _{\alpha
},\omega _{\alpha },H_{\alpha })$, where each $H_{\alpha }$ is a Hilbert
space of functions and $\omega _{\alpha }$ is an isomorphism of $\bf {S}$
onto $H_{\alpha }$ will be called a {\it functional atlas} on 
$\bf {S}$. A collection of all compatible functional atlaces on $\bf {S}
$ will be called a {\it coordinate structure} on $\bf {S}$.

{\it Definition}. Let $(U_{\alpha },\pi _{\alpha })$ be a chart on $\bf {S}$.
If $p\in U_{\alpha },$ then 
$\omega _{\alpha }\circ \pi _{\alpha }(p)$ is called the {\it coordinate} 
of $p$. The isomorphisms 
$\omega _{\beta }\circ \pi _{\beta }\circ (\omega _{\alpha }\circ 
\pi _{\alpha })^{-1}:\omega _{\alpha }\circ \pi _{\alpha }
(U_{\alpha }\cap U_{\beta })\longrightarrow 
\omega _{\beta }\circ \pi _{\beta }(U_{\alpha }\cap U_{\beta })$ are called
{\it coordinate transformations} on $\bf S$.

As $\bf {S}$ is a differentiable manifold one can also introduce the tangent
bundle structure $\tau :T{\bf{S}}\longrightarrow {\bf {S}}$ and the bundle $\tau
_{s}^{r}:T_{s}^{r}{\bf {S}}\longrightarrow {\bf {S}}$
of tensors of rank $(r,s)$.

\section{Coordinate transformations preserving locality of operators}
\setcounter{equation}{0}

In section 5 of \cite{Kryukov} it was shown that a single eigenvalue problem for an operator
on the string space leads to a family of eigenvalue problems in particular 
string bases. We now raise a more general question of describing changes of 
functional equations under transformations of string coordinates.
In particular, we will be interested in differential and 
algebraic equations.

When transforming a functional equation it is often necessary to 
preserve some of the properties of the equation. In particular, it is often useful
to preserve locality of the operators in the equation. To define locality 
of operators let us start with the following

{\it Definition.} A generalized function is {\it concentrated at a point} if 
it is equal to zero on every test function that is equal to zero on a neighborhood
of the point.

The structure of such functionals is given by the following theorem (see \cite{Gelfand1}):

{\it Theorem.} If the fundamental space contains all infinitely differentiable functions of
bounded support at least in some neighborhood of a given point $x_{0}$, then every generalized
function concentrated at $x_{0}$ has the form
\begin{equation}
\label{localfun}
f=\sum_{\left| q\right|\leq r}a_{q}D^{q}\delta(x-x_{0}).
\end{equation}
Here $x=(x_{1},...,x_{n})$ is a point in $R^{n}$, $r$ is a nonnegative integer, 
$q=(q_{1},...,q_{n})$ is a set of
nonnegative integers, $\left| q\right|=q_{1}+...+q_{n}$, and 
$D^{q}=\frac{\partial ^{|q|}}{\partial{x_{1}}^{q_{1}}...\partial{x_{n}} ^{q_{n}}}$.

{\it Definition.} Let $H$ be a coordinate space of functions on $R^{n}$. We shall say that a linear 
operator $A: H \longrightarrow H$ is local if
\begin{equation}
\label{local}
(Af)(x)=\int \sum_{\left| q\right|\leq r}a_{q}(x)D^{q}\delta(y-x)f(y)dy.
\end{equation}
Here $\delta(y-x)$ denotes the $\delta$-function of the diagonal $(x,x)$ in $R^{2n}$. 
Assume first that $f$ is an infinitely differentiable function of bounded support and
$a_{q}(x)D^{q}f(x)$ is integrable. 
Then formula (\ref{local}) is understood by requiring the validity of 
``integration by parts" which reduces (\ref{local}) to
\begin{equation}
\label{A}
(Af)(x)=\int \sum_{\left| q\right|\leq r}(-1)^{\left| q\right|}a_{q}(x)D^{q}f(x)dx,
\end{equation}
with integration over the entire space $R^{n}$. 
More generally, let $f \in H$ be any generalized function on $R^{n}$. Then (\ref{local})
is understood by requiring that
\begin{equation}
(Af,\varphi)=(f,B\varphi),
\end{equation}
where $\varphi$ is any smooth function of bounded support on $R^{n}$ and 
\begin{equation}
(B\varphi)(x)=\int \sum_{\left| q\right|\leq r}(-1)^{\left| q\right|}D^{q}(a_{q}(x)\varphi(x))dx.
\end{equation}
Here we assume that $a_{q}$ are smooth functions on $R^{n}$.

It is easy to see that locality of an operator is not an invariant property, that is, it 
depends on a particular choice of coordinates. We now want to describe such coordinate transformations
that preserve locality of linear operators.

Suppose then that $A:H \longrightarrow H$ is a local linear operator, 
$\omega: \widetilde{H} \longrightarrow H$ is a transformation of coordinates, and 
$A_{\widetilde{H}}=\omega^{-1}A\omega:\widetilde{H}\longrightarrow \widetilde{H}$ is the 
transformed operator.

The operator $A_{\widetilde{H}}$ will be local if
\begin{equation}
\sum_{\left| q\right|\leq r}\omega^{-1}(x,y)a_{q}(y)D^{q}\delta(z-y)\omega(z,u)f(u)dydzdu=
\sum_{\left| q\right|\leq s}b_{q}(x)D^{q}\delta(y-x)f(y)dy,
\end{equation}
where notations are as in (\ref{localfun}) and the integral symbol is omitted. That is,
\begin{equation}
\label{local1}
\sum_{\left| q\right|\leq r}a_{q}(x)D^{q}\delta(z-x)\omega(z,y)dz=
\sum_{\left| q\right|\leq s}\omega(x,z)b_{q}(z)D^{q}\delta(y-z)dz.
\end{equation}

In the simpliest case when $H$ and $\widetilde{H}$ are spaces of functions of one variable and 
the kernels of $A$ and $A_{\widetilde{H}}$ contain only one term
of the form $a_{q}(x)D^{q}\delta(y-x)$ each, the equation (\ref{local1}) reduces to
\begin{equation}
\label{local2}
a(x)\frac{\partial ^{n}}{\partial z^{n}}\delta(z-x)\omega(z,y)dz=
\omega(x,z)b(z)\frac{\partial ^{m}}{\partial y^{m}}\delta(y-z)dz.
\end{equation}
If particular, when $n=1$ and $m=0$ (\ref{local2}) yields
\begin{equation}
\label{local3}
a(x)\frac{\partial }{\partial z}\delta(z-x)\omega(z,y)dz=\omega(x,z)b(z)\delta(y-z)dz.
\end{equation}
Assuming that $\omega$ is a smooth solution, ``integration by parts" gives
\begin{equation}
\label{local10}
-a(x)\frac{\partial \omega(x,y)}{\partial x}=\omega(x,y)b(y).
\end{equation}
Solving (\ref{local10}) we obtain
\begin{equation}
\label{solution10}
\omega(x,y)=F(y)e^{-c(x)b(y)},
\end{equation}
where $c(x)=\int\frac{dy}{a(y)}$ and $F(y)$ is an arbitrary smooth function.
To be a coordinate transformation $\omega$ must be an isomorphism as well.
In particular, Fourier transform is a solution of (\ref{local10}) with 
\begin{equation}
\omega(x,y)=e^{ixy}.
\end{equation}

Coordinate transformations satisfying (\ref{local10}) preserve locality of the first order 
differential operators on $H$ by transforming them into operators of multiplication.  

In the case of a more general equation (\ref{local2}), we have:
\begin{equation}
\label{local-gen}
a(x)\frac{\partial^{n} \omega(x,y)}{\partial x^{n}}=
(-1)^{n}\frac{\partial^{m} (\omega(x,y)b(y))}{\partial y^{m}}.
\end{equation}
Solutions of (\ref{local-gen}) for different values of $n$ and $m$ produce 
coordinate transformations preserving locality of various differential operators.

\section{Coordinate transformations preserving derivatives}
\setcounter{equation}{0}

Among solutions of (\ref{local-gen}) those preserving the order $q$ of derivatives 
are of particular interest. To describe such transformations it is enough to obtain
solutions of (\ref{local-gen}) with $n=m=1$. Let us assume here that the coefficients $a$ and $b$ 
in (\ref{local-gen}) are constants. Then up to a constant coefficient which we assume to be 
equal to one we obtain the following equation:
\begin{equation}
\label{derivative}
\frac{\partial \omega(x,y)}{\partial x}+\frac{\partial \omega(x,y)}{\partial y}=0.
\end{equation}
The smooth solutions of (\ref{derivative}) are given by
\begin{equation}
\label{der-sol}
\omega(x,y)=f(x-y),
\end{equation}
where $f$ is an arbitrary infinitely differentiable function on $R$.
In particular, the function
\begin{equation}
\label{exp}
\omega(x,y)=e^{-(x-y)^{2}}
\end{equation}
satisfies (\ref{derivative}). Also, in section 4 of \cite{Kryukov} it was verified that the corresponding
transformation is injective. When Hilbert structure on $\widetilde{H}=\omega^{-1}(H)$
is induced by $\omega$, this transformation becomes an isomorphism of Hilbert spaces.
Therefore, it provides an example of a coordinate transformation that
preserves derivatives.

Let now $H$ be a Hilbert space of functions on $R^{n}$. 
We are looking for a nontrivial transformation preserving all partial derivative
operators on $H$. Applying equation (\ref{local1}) to this case we obtain:
\begin{equation}
\label{derivative1}
\frac{\partial \omega(x,y)}{\partial x_{i}}+\frac{\partial \omega(x,y)}{\partial y_{i}}=0.
\end{equation}
Here $i$ changes from $1$ to $n$. The function
\begin{equation}
\label{exp1}
\omega(x,y)=e^{-(x-y)^{2}}
\end{equation}
with $x=(x_{1},...,x_{n})$ and $y=(y_{1},...,y_{n})$  satisfies (\ref{derivative1}) and the
corresponding transformation is injective inducing a Hilbert structure on the space
$\widetilde{H}=\omega^{-1}(H)$.

{\it Theorem}. The generalized solutions of any linear differential equation with constant coefficients 
(either ordinary or partial) are coordinate transformations of the corresponding smooth solutions.
That is, let $L$ be a polynomial function of $n$ variables. Let $u, v \in \widetilde{H}$ be
functionals on the space $K$ of functions of $n$ variables which are infinitely differentiable 
and have bounded supports.
Assume that $u$ is a generalized solution of
\begin{equation}
L\left( \frac{\partial}{\partial{x}_{1}},...,\frac{\partial}{\partial{x}_{n}}\right)u=v.
\end{equation}
Then there exists a smooth solution $\varphi$ of 
\begin{equation}
L\left( \frac{\partial}{\partial{x}_{1}},...,\frac{\partial}{\partial{x}_{n}}\right)\varphi
=\psi,
\end{equation}
where $\varphi=\omega{u}$, $\psi=\omega{v}$ and $\omega$ is as in (\ref{exp1}).

{\it Proof}. Consider first the simpliest case of the ordinary differential equation
\begin{equation}
\label{diffeq}
\frac{d}{dx}u(x)=v(x).
\end{equation}
Assume $u$ is a generalized solution of (\ref{diffeq}). Define $\varphi=\omega{u}$ and
$\psi=\omega{v}$, where $\omega$ is as in (\ref{exp}). Notice that $\varphi, \psi$ are
infinitely differentiable. In fact, any functional on the space $K$ of infinitely
differentiable functions of bounded support acts as follows (see \cite{Gelfand1}):
\begin{equation}
(f,\varphi )=\int F(x)\varphi ^{(m)}(x)dx, 
\end{equation}
where $F$ is a continuous function on $R$. Applying $\omega$ to $f$ shows that the result
is a smooth function.

As $\omega^{-1}\frac{d}{dx}\omega=\frac{d}{dx}$, we have
\begin{equation}
\label{diffgen}
\omega^{-1}\frac{d}{dx}\omega{u}=v.
\end{equation}
That is,
\begin{equation}
\frac{d}{dx}\varphi(x)=\psi(x)
\end{equation}
proving the theorem in this case.
The higher order derivatives can be treated similarly as
\begin{equation}
\label{higherder}
\omega^{-1}\frac{d^n}{dx^{n}}\omega=
\omega^{-1}\frac{d}{dx}\omega \omega^{-1}\frac{d}{dx}\omega...
\omega^{-1}\frac{d}{dx}\omega.
\end{equation}
That is, transformation $\omega$ preserves derivatives of any order. 
Generalization to the case of several variables is straightforward.

\section{Coordinate transformations preserving products of functions}

\setcounter{equation}{0}

It is now natural to investigate changes of equations containing products
of functions under transformations of string coordinates.
Consider the simpliest algebraic equation
\begin{equation}
\label{product0}
a(x)f(x)=h(x),
\end{equation}
where $f$ is an unknown (generalized) function of a single variable and $h \in H$.
To investigate transformation properties of this equation we need to interpret
it as a tensor equation on the string space $\bf {S}$. The right hand side is 
a function. Therefore this must be a ``vector equation" (i.e. both sides must be $(1,0)$-tensors
on the string space).
If $f$ is to be a function as well, $a$ must  be a $(1,1)$-tensor. That is, the
``correct" equation is:
\begin{equation}
\label{product1}
a(x)\delta(x-y)f(y)dy=h(x).
\end{equation}
To preserve the product-like form of the equation we need such a coordinate transformation
$\omega:\widetilde{H}\longrightarrow H$ that
\begin{equation}
\label{product}
\omega^{-1}(u,x)a(x)\delta(x-y)\omega(y,z)dxdy=b(u)\delta(u-z).
\end{equation}
In this case the equation (\ref{product0}) in new coordinates is 
\begin{equation}
b(x)\varphi(x)=\psi(x),
\end{equation}
where $h=\omega{\psi}$, $f=\omega{\varphi}$, 
and $\varphi,\psi \in \widetilde{H}$. 

Equation (\ref{product}) is a particular case of equation (\ref{local-gen}) with $n=m=0$.
It yields
\begin{equation}
\label{aomega}
a(x)\omega(x,y)=\omega(x,y)b(y).
\end{equation}
If $a(x)=b(y)=C$, this equation is satisfied for any $\omega$. Otherwise 
$\omega$ must be a local transformation. 
The first case is trivial. In the second case we have 
\begin{equation}
\label{sum}
\omega(x,y)=\sum_{\left| q\right|\leq r}a_{q}(x)D^{q}\delta(y-x).
\end{equation}
Assume first that only one term in the sum (\ref{sum}) is present, i.e.
\begin{equation}
\omega(x,y)=a_{n}(x)D^{n}\delta(y-x).
\end{equation}
For now let us leave the question of invertibility of $\omega$ aside. 
Applying equation (\ref{aomega}) to a function $\varphi$ after ``integration by parts" we have
\begin{equation}
a(x)a_{n}(x)D^{n}\varphi(x)=a_{n}(x)D^{n}(b(x)\varphi(x)), 
\end{equation}
or,
\begin{equation}
\label{product2}
a(x)a_{n}(x)D^{n}\varphi(x)=a_{n}(x)\sum_{i+k=n}\frac{k!}{i!(k-i)!}D^{i}b(x)D^{k}\varphi(x)).
\end{equation}

Choose $\varphi$ so that the derivatives $\varphi^{(k)}$ form a (classical) basis on 
$\widetilde{H}$. Then by equating the 
coefficients of $\varphi^{(n)}$ we have $a(x)=b(x)$. If $n>0$, we also have $b'(x)=0$. That is, 
if $n>0$, then $a(x)=b(x)=C$ in which case $\omega$ can be any.
If $n=0$ instead, then $a(x)=b(x)$ can be any. In this case, however,
\begin{equation}
\omega(x,y)=a_{0}(x)\delta(y-x),
\end{equation}
i.e. the transformation is simply multiplication by a function. 

In more general case when $\omega$ is as in (\ref{sum}) a similar analysis gives the same
result: whether $a(x)=b(x)=C$, or $\omega(x,y)=a_{0}(x)\delta(x-y)$. 

We therefore have the following

{\it Theorem}. Unless $a(x)$ is a constant or $\omega(x,y)=a_{0}(x)\delta(x-y)$, 
it is impossible to preserve the product
form of (\ref{product1}) under coordinate transformations. 

In particular, the product of nonconstant functions of one and the same variable is not 
an invariant operation.

One could refer to the operator $a(x)\delta(x-y)$ in (\ref{product1}) as the 
operator of multiplication by $a(x)$. Clearly, it is a local operator. 
The theorem then says that locality of this operator can be preserved
only in trivial cases when $a(x)=C$ or $\omega$ itself is an operator of
multiplication by a function.

On the other hand,  consider the equation
\begin{equation}
a(x)f(y)=h(x,y),
\end{equation}
where $a$ and $f$ are functions of a single variable and $h$ is a function of two variables.
This equation can be viewed as a tensor equation on the string space. The left hand side
represents then a tensor product of two ``vectors". The right hand side is a $(2,0)$-tensor.
Therefore, any coordinate transformation preserves this form of the equation. In particular,
generalized solutions of this equation can be transformed into ordinary solutions by 
transformation of coordinates. 

\section{More general coordinate transformations}
\setcounter{equation}{0}

In section 4 we have studied coordinate transformations preserving linear differential operators 
with constant coefficients. Here we will investigate  the case of linear
differential operators with non-constant coefficients. We will also begin analyzing coordinate
transformations of nonlinear differential equations. We start with the following

{\it Example}. Consider the equation (\ref{local-gen}) with $n=m=1$ assuming $a(x)$ and $b(y)$ are functions.
In this case the equation reads
\begin{equation}
\label{nonconst}
a(x)\frac{\partial{\omega(x,y)}}{\partial{x}}+\frac{\partial{(\omega(x,y)b(y))}}{\partial{y}}=0.
\end{equation}
Let us look for a solution in the form
\begin{equation}
\omega(x,y)=e^{f(x)g(y)}.
\end{equation}
Then (\ref{nonconst}) yields
\begin{equation}
\label{nonconst1}
a(x)f'(x)g(y)+b(y)f(x)g'(y)+b'(y)=0.
\end{equation}
If $b(y)=1$, (\ref{nonconst1}) is a separable equation and we have
\begin{equation}
\frac{a(x)f^{\prime }(x)}{f(x)}=-\frac{g^{\prime }(y)}{g(y)}=C,
\end{equation}
where $C$ is a constant. Solving this we have,
\begin{equation}
\omega(x,y)=e^{Ce^{\int \frac{C_{1}}{a(x)}dx} e^{C_{2}y}}.
\end{equation}
Taking for example $C=C_{1}=C_{2}=1$ and $a(x)=x$, we have
\begin{equation}
\omega(x,y)=e^{xe^{y}}.
\end{equation}
The corresponding transformation can be shown to be invertible on an appropriate space of functions. 
As we see it transforms the operator $x\delta^{\prime}(y-x)$ into the operator $\delta^{\prime}(y-x)$. That is, 
\begin{equation}
\omega: x \psi^{\prime}(x) \longrightarrow \psi^{\prime}(x)
\end{equation}
for {\it any} function $\psi$ on the space of definition of $\omega$.

{\it Example}. As another example consider (\ref{local-gen}) with $n=2,m=0$.
We have:
\begin{equation}
a(x)\frac{\partial^{2}{\omega(x,y)}}{\partial{x^{2}}}=\omega(x,y)b(y)
\end{equation}
Looking for a solution in the form $\omega(x,y)=e^{f(x,y)}$ we obtain:
\begin{equation}
f_{xx}+f_{x}^{2}=\frac{b(y)}{a(x)},
\end{equation}
where $f_{x}, f_{xx}$ denote partial derivatives of $f(x,y)$ with respect to $x$. 
Using $g(x,y)=f_{x}(x,y)$ we obtain a first order differential equation
\begin{equation}
\label{nonlin}
g_{x}+g^{2}=\frac{b(y)}{a(x)}.
\end{equation}
In particular,
when $b(y)=y^{2}$ and $a(x)=1$ we are back to the Fourier-like transform as in (\ref{solution10}).

We see from the previous examples that in solving equation (\ref{local-gen}) we need to take into account the
specifics of a problem in hand. A type of coordinate transformation especially useful to
treat the problem is determined by a type of problem itself.

A very important question is whether we can apply the developed coordinate formalism to nonlinear differential 
equations. It is known that the theory of generalized functions has been mainly successful with the 
linear problems. The difficulty of corse lies in defining the product of generalized functions.
To see what kind of solution can be offered in the new context consider the following

{\it Example}. Consider a differential equation containing the square of derivative of an unknown function,
i.e. containing the term 
\begin{equation}
\label{nonlin0}
\varphi^{\prime}(x)\cdot \varphi^{\prime}(x),
\end{equation}
where $\varphi \in H$.
To use the coordinate formalism we need to interpret this term as a tensor. We have:
\begin{equation}
\label{nonlin1}
\varphi^{\prime}(x)\cdot \varphi^{\prime}(x)=
\delta(x-y)\delta^{\prime}(u-x)\delta^{\prime}(v-y)\varphi(u)\varphi(v)dydudv,
\end{equation}
where as before we omit the integral symbol.
Therefore, this term is the convolution of the $(1,2)$-tensor
\begin{equation}
\label{nonlin2}
c^{x}_{uv}=\delta(x-y)\delta^{\prime}(u-x)\delta^{\prime}(v-y)dy
\end{equation}
with the pair of strings $\varphi^{u}=\varphi(u)$.
With this interpretation we can easily obtain the transformation properties of $c^{x}_{uv}$. Denote
\begin{equation}
\label{nonlin3}
c^{x}_{uv}\varphi^{u}\varphi^{v}=\psi^{x},
\end{equation}
where the meaning of notations is described in section 2.
Assume now that $\omega: {\widetilde{H}}\longrightarrow H$ is a coordinate transformation and 
$\omega {\widetilde{\varphi}}= \varphi$. Denote $\omega(x,y)=\omega^{x}_{y}$.
As $\psi^{x}$ is a ``vector", we have 
\begin{equation}
\label{nonlin4}
c^{x}_{uv}\omega^{u}_{u^{\prime}}\widetilde{\varphi^{u^{\prime}}}\omega^{v}_{v^{\prime}}\widetilde{\varphi^{v^{\prime}}}
=\omega^{x}_{x^{\prime}}\widetilde{\psi^{x^{\prime}}}.
\end{equation}
That is,
\begin{equation}
\label{nonlin5}
c^{x^{\prime}}_{u^{\prime}v^{\prime}}=
{\omega^{-1}}^{x^{\prime}}_{x}{\omega^{*}}^{v}_{v^{\prime}}c^{x}_{uv}\omega^{u}_{u^{\prime}},
\end{equation}
where ${\omega^{*}}$ is the adjoint of $\omega$.
After ``integration by parts"  we obtain
\begin{equation}
\label{nonlin6}
c^{x}_{u^{\prime}v^{\prime}}=
{\omega^{*}}^{v}_{v^{\prime}}c^{x}_{uv}\omega^{u}_{u^{\prime}}=
\frac{\partial{\omega(v^{\prime},x)}}{\partial{x}}\frac{\partial{\omega(x,u^{\prime})}}{\partial{x}}.
\end{equation}
By specifying the desired form of $c^{x^{\prime}}_{u^{\prime}v^{\prime}}$ we obtain a nonlinear partial
differential equation for the transformation $\omega$.
Existence of interesting solutions of this and similar equations is under investigation.

\section{Conclusion}
\setcounter{equation}{0}

The main idea of the coordinate formalism introduced in \cite{Kryukov} is to relate different spaces of 
functions by considering them as coordinate representations of an invariant ``string" space. This last one 
is simply the abstract infinite-dimensional separable Hilbert space. This approach turns out to be very 
similar in spirit to the nineteenth century introduction of vectors. However, it can {\it not} be reduced to
consideration of elements of an infinite-dimensional Hilbert space as vectors. In fact, given a Hilbert
space of functions elements of such space {\it are} vectors. The objects that we call strings are more general.
They are defined for all Hilbert spaces of functions (i.e. coordinate spaces) at once and do not 
depend on a choice of such space. 

A particular choice of a coordinate space can be useful for a problem in hand. Therefore, special transformations
of coordinates become important. In particular, transformations from the spaces of ordinary functions to the spaces
of generalized functions provide a new insight on the theory of generalized functions.

Here we saw how preservation of different properties of linear operators led to different types of coordinate
transformations. For example, preserving locality of operators in the simpliest case leads to the Fourier-like 
transformations as in (\ref{solution10}). 

By requiring preservation of the derivative operator we were able to relate the generalized and the smooth solutions 
to linear partial differential equations with constant coefficients.

The results of the last section suggest that the nonlinear problems can be approached in the same fashion. For this
it must be possible to interpret a given nonlinear equation as a tensor equation on the string space. Then 
different ``nonlinearities" are interpreted as convolutions of tensors on the space. A more complete 
analysis in this direction is, however, a subject for a different paper.

%%%%%%%%%%%%%%%%%

\end{document}